\RequirePackage{lineno} 
\documentclass[aps,twocolumn,groupedaddress]{revtex4}
\usepackage{graphicx}  
\usepackage{dcolumn}   
\usepackage{bm}        
\usepackage{amssymb}   
\usepackage{multirow}
\usepackage{endnotes}
\usepackage{amsmath, amsthm}
\usepackage[all]{xy}
\usepackage{verbatim}
\usepackage{color}
\usepackage{lineno}
\usepackage{array,hhline,dcolumn} 
\usepackage[nice]{nicefrac}

\def\th{\theta}

\def\to{\rightarrow}

\def\beq{\begin{eqnarray}}
\def\eeq{\end{eqnarray}}
\def\fr#1#2{{{#1} \over {#2}}}

\def\numu{\nu_\mu}
\def\numub{\bar{\nu}_\mu}

\begin{document}

\title{Discrimination of parametrizations for nuclear effects in neutrino scattering through comparisons of low ($\sim$ 700 MeV) and medium ($\sim$ 3 GeV) energy cross-section data}
\date{\today}
\author{
		J. Grange$^1$\footnote{Present address: Argonne National Laboratory, Argonne, IL, 60439, U.S.A}, C. Juszczak$^2$, J. Sobczyk$^2$ and G. P. Zeller$^3$  \\
}
\smallskip
\smallskip
\affiliation{
$^1$University of Florida; Gainesville, FL 32611 \\
$^2$Institute of Theoretical Physics, Wroc\l aw University; Wroc\l aw, Poland \\
$^3$Fermi National Accelerator Laboratory; Batavia, IL 60510 \\
}

\begin{abstract}
High-quality charged current quasielastic scattering data have recently been reported for both muon 
neutrinos and antineutrinos from several accelerator-based neutrino experiments. Measurements from 
MiniBooNE were the first to indicate that more complex nuclear effects, now thought to be the result 
of nucleon pair correlations, may contribute to neutrino quasielastic samples at a much higher 
significance than previously assumed. These findings are now being tested by MINER$\nu$A and other 
contemporary neutrino experiments. Presented here is a comparison of data from MiniBooNE and 
MINER$\nu$A to a few example parametrizations of these nuclear effects. It has been demonstrated 
that such effects may bias future measurements of neutrino oscillation parameters and so this 
issue continues to press the neutrino community. A comparison of data over a large range of 
neutrino energies is one approach to exploring the extent to which such nucleon correlations may 
influence our understanding and subsequent modeling of neutrino quasielastic scattering.
\end{abstract}

\maketitle

Charged current quasielastic scattering (CCQE, $\nu_l + N \to l + N'$) is the 
dominant interaction channel in many neutrino oscillation measurements.
In practice, the assumed simple multiplicity and topology of such processes 
allow for the recovery of the incident neutrino energy (an essential quantity 
in neutrino oscillation fits) using only measurement of the outgoing charged 
lepton. Assuming background processes can be reliably subtracted with adequate 
precision, such CCQE samples then become an attractive channel through which to 
extract oscillation parameters because the sole reliance on lepton kinematics avoids
experimental complications associated with the need to explicitly reconstruct final state nucleons.

Recently, there has been mounting evidence to suggest that such CCQE processes 
may not be as simple as originally thought, particularly when scattering 
off nuclear targets~\cite{qeReview}. The presence of correlated nucleon pairs 
in the nuclear environment may alter both the magnitude and kinematics of these 
interactions at a significant level. Resultant enhancements have been previously 
observed in transverse electron-nucleus data~\cite{carlson}, but the role such 
effects play is only now being appreciated in the context of neutrino-nucleus 
scattering, in large part motivated by the MiniBooNE data~\cite{mbNu,mbNub}.
 
Of course, it is important to get the physics right. The complex nuclear environment 
can have a potentially large impact on the determination of neutrino energy, 
(anti)neutrino rates, and nucleon emission in neutrino oscillation 
analyses~\cite{meloni,meloniMartT2k,huberOsc,martiniOsc,moselEnuReco}. 
Additionally, some amount of model dependence enters the cross-section data
through the necessary reliance (to some degree) upon an event generator for the purpose of 
background prediction and subtraction.  Hence, much attention has
been devoted to this topic in recent years. While the theoretical and experimental 
understanding of this issue is still taking shape, most neutrino experiments do not 
currently include a complete implementation of nuclear effects (including nucleon 
correlations) in their simulations. Lacking this, confrontation of the experimental 
data and leading models have often been limited to comparisons of the absolute 
cross section as a function of neutrino energy, $E_{\nu}$, and hence suffer 
from model dependences inherent in extracting $E_{\nu}$ from the data. High 
statistics information from MiniBooNE has recently changed this and allowed 
detailed comparison of nuclear models to flux-averaged double differential 
distributions of the observed muon kinematics, available for the first time for both neutrino 
and antineutrino quasielastic scattering on carbon~\cite{mbNu,mbNub}. Furthermore, 
the full angular coverage of the final state muon offered by the spherically-symmetric 
detector allows a unique test of the transverse enhancement expected due to nucleon pair 
correlations (such effects are expected to be largest for backwards-scattered muons relative
to the incoming neutrino beam). 

More recently, MINER$\nu$A has reported measurements of the flux-averaged
differential cross section, $d\sigma/dQ^2_{QE}$, for both neutrino and 
antineutrino quasielastic scattering also on a carbon-based target~\cite{minNu,minNub}. 
The analysis of the MINER$\nu$A data further includes an exploitation of 
it's fine-grained calorimetry to scrutinize hadronic activity near the 
quasielastic interaction vertex. Like the earlier MiniBooNE findings, the 
results suggest the presence of nuclear effects not included in widely-used 
relativistic Fermi Gas (RFG)~\cite{RFG} models which assume independent (and not correlated)
nucleons in the nucleus. To facilitate a more direct comparison of the MiniBooNE 
and MINER$\nu$A data, we present a recasting of the MiniBooNE experimental data 
in the same form as recently reported by MINER$\nu$A~\cite{minQsqPlots}. 

Here, the exercise of producing normalized ratios in $Q^2_{QE}$ with respect 
to the nominal RFG model, as presented by MINER$\nu$A, is repeated with the 
MiniBooNE neutrino and antineutrino quasielastic data. In this case, 
$Q^2_{QE}$ refers to the squared four-momentum transfer obtained using 
only reconstructed muon kinematics and assuming quasielastic scattering 
with a single target nucleon at rest:

\beq
E_\nu^{QE}
&=&
\fr{2(M_n^{\prime})E_\mu-((M_n^{\prime})^2+m_\mu^2-M_p^2)}
{2\cdot[(M_n^{\prime})-E_\mu+\sqrt{E_\mu^2-m_\mu^2}\cos\th_\mu]},\label{eq:recEnu}\\
Q^2_{QE}
&=&
-m_\mu^2+2E_\nu^{QE}(E_\mu-\sqrt{E_\mu^2-m_\mu^2}\cos\th_\mu),\label{eq:recQsq}
\eeq

\noindent
where $E_\mu = T_\mu + m_\mu$ is the total muon energy and $M_n$, $M_p$, 
$m_\mu$ are the neutron, proton, and muon masses. The adjusted neutron mass, 
$M_n^{\prime}=M_n-E_B$, depends on the separation energy in carbon, $E_B$, 
which is set to 34 (30)~MeV for neutrino (antineutrino) scattering.  
Note that the $Q^2_{QE}$ formula explicitly assumes one-body QE interactions.  
While this assumption may be faulty, the comparison is well justified because the 
prediction also assumes the same condition.  Therefore, we are able to compare 
exactly the same observable quantity and learn from the level of consistency. 
The deviation from true Q$^2$ is present in the experimental data due to nuclear 
effects and the inclusion of two body current contributions are implemented into 
the prediction. Divergences between the prediction and data show that Monte Carlo
models are not perfect, but the comparison remains meaningful.

\begin{figure}[h]
\begin{center}
\includegraphics[scale=0.46]{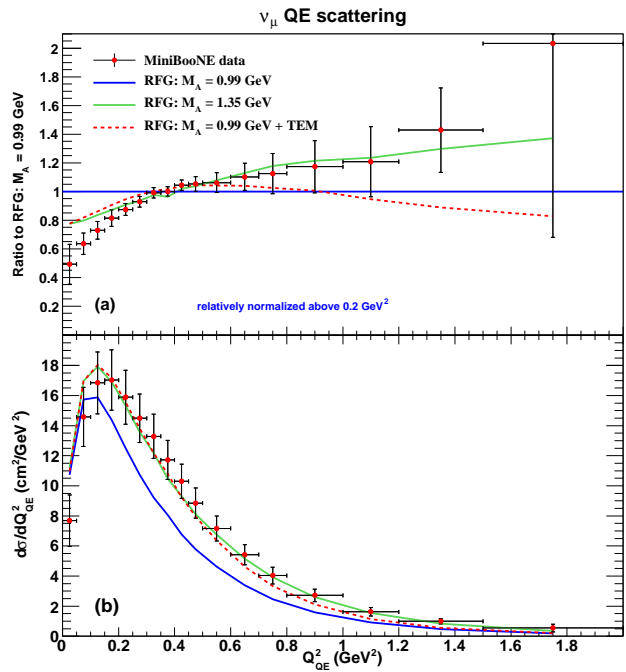} 
\end{center} 
\caption{(Color online) The shape (a) and scale (b) of $\numu$ MiniBooNE 
$Q^2_{QE}$ data compared to parametrizations of the RFG presented in the same 
form as MINER$\nu$A data in recent publications~\cite{minNu,minNub,minQsqPlots}. 
Appropriate to each comparison, shape-only uncertainties accompany the data in 
(a), while total uncertainties are shown in (b).  Within these experimental uncertainties, 
in the MiniBooNE energy range the effect of treating nuclear effects with an increase in the axial mass is 
largely consistent with the TEM description both in shape and in scale.  
Note that Pauli blocking has not been tuned in the models shown here, and so 
the agreement in the low Q$^2_{QE}$ region is somewhat worse compared to the tuned distributions 
shown in Refs.~\cite{mbNu} and~\cite{mbNub}.}
\label{fig:qsqNu}
\end{figure}

\begin{figure}[h]
\begin{center}
\includegraphics[scale=0.46]{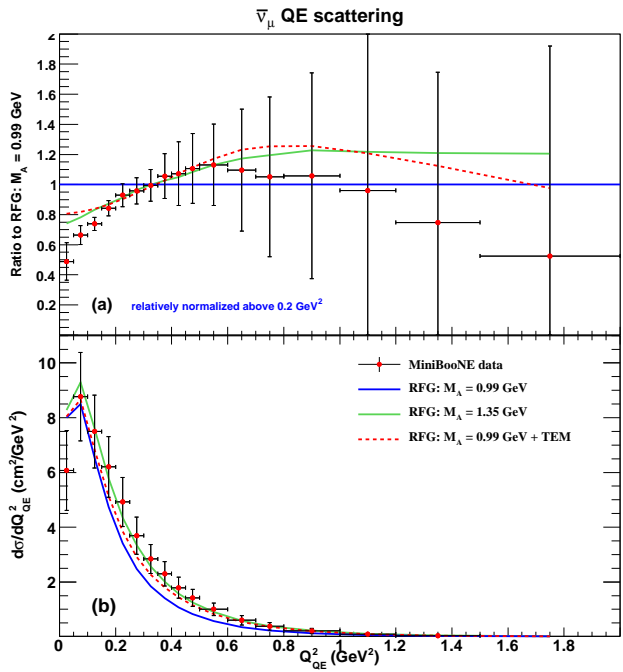} 
\end{center} 
\caption{(Color online) The same distributions described in Figure~\ref{fig:qsqNu}, but for $\numub$.  Note that, as is also the case for $\numu$, the $\numub$ data appears simultaneously consistent with an increase in the axial mass and the introduction of the TEM.}
\label{fig:qsqNub}
\end{figure}

The results, along with the bare $Q^2_{QE}$ distributions used to produce the 
shape comparison, are presented for neutrinos
in Figure~\ref{fig:qsqNu} and for antineutrinos in Figure~\ref{fig:qsqNub}. In 
following what was reported by MINER$\nu$A~\cite{minNu,minNub,minQsqPlots}, a comparison is shown 
for two example alternatives: increasing the axial mass parameter, $M_A$, in the 
RFG model and including a parametrization of the transverse enhancement seen in 
electron-nucleus scattering. Both have been motivated by the MiniBooNE observations 
and are shown to provide viable descriptions of this data. The value for $M_A$ is 
chosen from spectral fits to the MiniBooNE CCQE events~\cite{mbNu} while electron 
scattering data on heavy nuclei provide the formulation of the Transverse 
Enhancement Model (TEM). In this implementation, the TEM specifically modifies 
the magnetic form factor for bound nucleons to achieve simultaneous agreement both 
with a wide range of electron scattering data and the early neutrino cross-section 
measurements on deuterium~\cite{bodek}. In the absence of a full nuclear physics
description in neutrino event generators, such parametrizations can be a helpful 
tool for testing the gross features of such contributions and comparing data sets.

As for all models based on the impulse approximation, precision is not expected 
from the RFG in the region of small $Q^2$~\cite{lowQsq}.  For this reason, we focus 
on the higher $Q^2$ data and normalize the distributions presented here using the region 
$Q^2_{QE}>0.2$ GeV$^2$.   This excludes the most uncertain region of momentum transfer, $q<$450 MeV. The inclusion of RPA
effects increases the accuracy of predictions in this region,
and {\it are not} included in the models presented here.
The ratios of the various distributions are executed after requiring the area of 
each differential cross section above 0.2~GeV$^2$ to match the cross-section strength in the
same region of the the nominal RFG model with $M_A$ = 0.99~GeV.  Apart from the focus on 
 $Q^2_{QE} > 0.2$ GeV$^2$, the philosophy of these shape comparisons is identical to that 
 of the MINERVA analyses.

To mitigate possible misinterpretations of these results 
due to the use of different neutrino interaction generators, the parametrizations 
of nuclear effects shown here are provided by the same NuWro neutrino event 
generator~\cite{nuWro} as used in the published MINER$\nu$A results.


Both increases to the axial mass and the inclusion of transverse enhancement 
effects have been suggested as options to describe the MiniBooNE data. As 
seen in Figures~\ref{fig:qsqNu} and~\ref{fig:qsqNub}, in general these adjustments seem to perform equally 
well. There seems to be some mild tension at high $Q^2_{QE}$ between the TEM 
and the MiniBooNE $\numu$ data; this has also been observed in comparisons of the 
model-independent double-differential distribution~\cite{jan}.  Because the inclusion of enhancements to 
the transverse current and increasing the axial mass can lead to similar 
results in the kinematic region accessed by the MiniBooNE flux, it can be difficult to disentangle their 
contributions.   Future
high-precision neutrino scattering experiments sensitive to this region such as 
MicroBooNE~\cite{uB}, NO$\nu$A~\cite{nova} and T2K~\cite{t2k} may be able to provide a more discriminatory test 
of high momentum transfer interactions.  Note that the MINER$\nu$A requirement of matching muons in the downstream calorimeter in order to recover the charge and momentum imposes an effective Q$^2$ cut on the analysis sample.  It may
be possible to extend the kinematic range accessed by implementing the kinds of techniques described in Refs.~\cite{mbNub,MBWS}.

When confronting these MiniBooNE plots with the similar 
version from MINER$\nu$A~\cite{minQsqPlots}, the benefit of comparing data sets 
across very different neutrino energy ranges is immediately apparent. 
While the changes associated with an increase in the axial mass and the 
inclusion of transverse enhancement effects (according to the TEM) have very 
similar effects at low MiniBooNE energies ($0.4 < E_\nu < 2$ GeV), the 
differences are much larger for higher MINER$\nu$A energies ($1.5 < E_\nu < 10$ GeV) 
where the two effects start to pull apart. In the case of MINER$\nu$A, 
a large increase in $M_A$ is not supported by the data and the TEM is more 
strongly favored~\cite{minQsqPlots}. Separating such nuclear effects from 
changes to the axial-vector form factor is important given that the two 
choices have very different implications for the interpretation of neutrino 
oscillation data. 

The recent reports of the MINER$\nu$A and MiniBooNE 
CCQE data significantly extend the experimental knowledge of neutrino and 
antineutrino interactions on carbon nuclei.  This robust collection of data 
offers an opportunity to directly test parametrizations of nuclear effects 
with neutrinos and antineutrinos across energy regimes crucial for current 
and next-generation oscillation experiments.  

It will be interesting to repeat similar cross-comparisons with more 
sophisticated nuclear models such as microscopic calculations of
multi-nucleon knock-out mechanisms~\cite{mart,nieves}.   An issue common 
to many such models is that they are reliable for the region of four-momentum transfer
dominantly accessed by the MiniBooNE neutrino flux but not for the MINERvA flux~\cite{gran}. 
Moreover, implementation of such models in Monte Carlo simulation requires the consistent inclusion of RPA 
effects, which leads to a considerably more complicated simulation scheme 
compared to present designs.  We emphasize again that the model parameterizations 
compared to the MiniBooNE data in Figures~\ref{fig:qsqNu},~\ref{fig:qsqNub} and the MINER$\nu$A data in Refs.~\cite{minNu} and~\cite{minNub} are limited in scope.  While more realistic prescriptions are becoming available in the literature, the somewhat naive models discussed here are likely to see continued use in neutrino experiments.  For this reason, their success in describing historical data sets is important to track and may be used to identify which features perform well in the context of the neutrino energies and kinematic regions accessed by unique experiments.

It will be with the sorts of high-resolution observations of both leptonic and 
hadronic activity in CCQE-like interactions presented by MiniBooNE and MINER$\nu$A, 
along with model-independent comparisons 
such as those presented here, that the neutrino interaction community will arrive 
at a definitive resolution to the size and kinematics of these important nuclear 
effects and what remaining role the axial-vector form factor plays.  New data and 
improved analyses from the MiniBooNE, MINER$\nu$A, SciBooNE~\cite{SB}, 
MicroBooNE~\cite{uB}, ArgoNeuT~\cite{argoneut}, ICARUS~\cite{icarus}, 
NOMAD~\cite{nomad} and the near detectors of the T2K~\cite{t2k}, NO$\nu$A~\cite{nova} 
and MINOS~\cite{minos} experiments are expected to play vital roles in this campaign.  
Meanwhile, the continued aggressive theoretical progress and anticipated integration 
into neutrino generators used by experiments will be invaluable towards understanding 
the fundamental basis for these interactions. 

C.J. and J.T.S. were partially supported by the grant 4574/PB/IFT/12 (UMO-2011/01/M/ST2/02578).


\begin{thebibliography}{99}

\bibitem{qeReview}
H. Gallagher, G. Garvey and G. P. Zeller,
Ann. Rev. Nucl. Sci. {\bf 61}, 355 (2011).

\bibitem{carlson}
J. Carlson {it et al.}, Phys. Rev. {\bf C65}, 024002 (2002).

\bibitem{mbNu}
A. A. Aguilar-Arevalo {\sl et al.} (MiniBooNE Collaboration), 
Phys. Rev. {\bf D81}, 092005 (2010).

\bibitem{mbNub}
A. A. Aguilar-Arevalo {\sl et al.} (MiniBooNE Collaboration), 
Phys. Rev. {\bf D88}, 032001 (2013).

\bibitem{meloni}
D. Meloni, J. Phys. Conf. Ser. {\bf 408}, 012024 (2013).

\bibitem{meloniMartT2k}
D. Meloni and M. Martini,
Phys. Lett. {\bf B716}, 186 (2012).

\bibitem{huberOsc}
P. Coloma, P. Huber, C. M. Jen and C. Mariani.
arxiv:1311.4506 [hep-ph].

\bibitem{martiniOsc}
M. Martini {\it et al.},
arXiv:1202.4745 [hep-ph].

\bibitem{moselEnuReco}
O. Lalakulich and U. Mosel,
Phys. Rev. {\bf C86}, 054606 (2012).

\bibitem{minNu}
G. A. Fiorentini {\sl et al.} (MINER$\nu$A Collaboration),
Phys. Rev. Lett. {\bf 111}, 022502 (2013).

\bibitem{minNub}
L. Fields {\sl et al.} (MINER$\nu$A Collaboration),
Phys. Rev. Lett. {\bf 111}, 022501 (2013).

\bibitem{RFG} 
R.~A.~Smith and E.~J.~Moniz, 
Nucl. Phys. {\bf B43}, 605 (1972);
{\it erratum: ibid.} {\bf B101}, 547 (1975).

\bibitem{minQsqPlots}
D. Schmitz on behalf of the MINER$\nu$A Collaboration,
``Quasi-Elastic Scattering of Neutrinos and Anti-Neutrinos at MINERvA".
Retreived from
http://theory.fnal.gov/jetp/talks/Schmitz\_WandC\newline\_MinervaCCQE\_2013\_05\_10.pdf

\bibitem{bodek}
A. Bodek, H. Budd and M.E. Christy. 
Eur. Phys. J. {\bf C71}, 1726 (2011).

\bibitem{lowQsq}
A. Ankowski, O. Benhar and N. Farina,
Phys. Rev. {\bf D82}, 013002 (2010).

\bibitem{nuWro}
T. Golan, C. Juszczak and J. Sobczyk,
Phys. Rev. {\bf C86}, 015505 (2012).

\bibitem{jan}
J.T. Sobczyk, Eur. Phys. J. {\bf C72}, 1850 (2012).








\bibitem{uB}
M. Soderberg for the MicroBooNE Collaboration, AIP Conf.
Proc. 1189, 83 (2009).

\bibitem{nova}
D.~S.~Ayres {\it et al.}  (NOvA Collaboration),
FERMILAB-DESIGN-2007-01. 

\bibitem{t2k}
K. Abe {\it et al.} (T2K Collaboration),
Phys. Rev. {\bf D88}, 032002 (2013).

\bibitem{MBWS}
A. A. Aguilar-Arevalo {\sl et al.} (MiniBooNE Collaboration), 
Phys. Rev. {\bf D84}, 072005 (2011). 

\bibitem{mart}
M. Martini, M. Ericson, G. Chanfray and J. Marteau, 
Phys. Rev. {\bf C80}, 065501 (2009).

\bibitem{nieves}
J. Nieves, I. Ruiz Simo and M.J. Vicente Vacas,
Phys. Rev. {\bf C83}, 045501 (2011).

\bibitem{gran}
R. Gran, J.Nieves, F. Sanchez and M.J. Vicente Vacas,
arXiv:1307.8105.

\bibitem{SB}
Y. Nakajima {\it et al.} (SciBooNE Collaboration),
Phys. Rev. {\bf D83}, 012005 (2011).  
  
\bibitem{argoneut}  
C. Anderson {\it et al.} (ArgoNeuT Collaboration),
Phys. Rev. Lett. {\bf 108}, 161802 (2012).

\bibitem{icarus}  
S. Amoruso {\it et al.} (ICARUS Collaboration),
Eur. Phys. J. {\bf C33}, 233 (2004).

\bibitem{nomad}
V. Lyubushkin {\it et al.} (NOMAD Collaboration),
Eur. Phys. J. {\bf C63}, 355 (2009).

\bibitem{minos}
P. Adamson {\it et al.} (MINOS Collaboration),
Phys. Rev. {\bf D81}, 072002 (2010).

\end{thebibliography}
\end{document}